\documentstyle[twocolumn,pra,aps]{revtex}

\tighten

\begin{document}
\title{A dynamical model of the chemical bond: \\
kinetic energy resonances between atomic orbitals}
\author{Holger F. Hofmann\\ German Aerospace Research Establishment\\
Institute of Technical Physics\\ Pfaffenwaldring 38-40\\ 70569 Stuttgart\\ 
Germany}
\maketitle
\begin {abstract}
A new approach to chemical bonding is introduced in order to 
provide an improved understanding of the connection between basic quantum 
mechanics and the covalent pair bond.
It's focus is on the fact that the energy of the bond is
largely given by the kinetic energy of the electrons, while the Coulomb
forces are only comparable to the kinetic energy terms close to the 
atomic nuclei, where they define the shape and the size of the atomic
orbitals. If atomic orbitals are used as a starting point, the kinetic
energy operator is sufficient to determine the energy of the chemical bond.
The simple mathematical structure of this operator allows a calculation of
bond energy as a function of the distance between the atoms. For Gaussian 
wavefunctions, it is
possible to calculate this bonding potential analytically, determining the
bond length, the bond energy and the elasticity of the bond from only a
single parameter, the width of the atomic wavefunction. It is shown
that the results correspond surprisingly well with experimental values for
diatomic molecules.
\end{abstract}

\section{Introduction}
While the theory of chemical bonding has advanced rapidly as more and
more details of the quantum mechanical problem have been taken into
account, surprisingly little effort has gone into the developement of
intuitive models to interpret the results obtained by more sophisticated
means. As a result, the theory of electronic structure is still introduced 
quite differently
to chemists (for example \cite{1}) and to physisists (for example \cite{2}). 
While the former begin with energetic considerations
in small molecules, the latter tend to focus on dynamics and solid state 
systems. Yet, the nature of the bonds is quite similar in
solids and in molecules, and a lot can be learned by examining common
features. 

The Sommerfeld theory of electrons in metals started out by completely
ignoring the Coulomb interactions of electrons and nuclei, treating the 
electrons as free particles moving in an effective vacuum. At first sight,
this free electron nature of the metallic bond seems to be qualitatively
different from covalent bonds.
However, in 1979 Froyen and Harrison \cite{3} showed, that the parameters 
for
the LCAO or tight binding approach to the band structure of tetrahedrally 
bound
solids such as Diamond, Silicon and Germanium can be derived from a
comparison with free electrons. Consequently, the matrix elements of
the bonding overlap depend only on the kinetic energy.
The potential energy enters through the separation of the atomic
energy levels and through the bond length, both of which are features of 
the atomic wavefunction.

It is therefore possible to explain bonding between atoms using only
the contribution of the kinetic energy to the bond formation, while the
Coulomb energy is responsible for tightly binding the electrons to the atoms.
This qualitative difference between the energy of atomic orbitals
and the energy of molecular bonds is crucial for understanding the 
physics of the chemical bond.

\section{Mechanical Properties of single Atoms: 
A simple Approach to Understanding Compressibility}
Electrons are fermions. They obey the Pauli principle. Therefore, 
two electrons of the same spin can only occupy the same region of space 
if they have different momentum. As a result, an increase in electron 
density requires an increase in kinetic energy and therefore electrons
resist compression even if we neglect the Coulomb repulsion. In fact,
for high densities, the Pauli principle turns out to be a stronger force
of repulsion than the Coulomb forces. 

The correct quantitative expression can be derived
from the Sommerfeld quantization rule that the density of quantum states 
in phase space is $h^{-3}$. The electron spin doubles this density of states,
as there are two quantized spin directions for each phase space volume
of $h^3$. The minimal energy for a given density of electrons may
now be determined by filling the lowest energy regions of phase space
and calculating the average kinetic energy $E_{Pauli}$ of each electron 
as a function of the electron density $\rho$. 
The result, usually derived as part of the Sommerfeld theory of electrons 
in metals (for example in chapter 2 of \cite{2}), is:

\begin{equation}
E_{Pauli}=\frac{3\hbar^2}{10m}(3\pi^2 \rho)^{2/3}
\end{equation}

where m is the electron mass and $\rho$ is the electron density.

This term, which increases with decreasing distance as $r^{-2}$, 
prevents the Coulomb attraction, which increases only with $r^{-1}$, 
from pulling the electrons into the
atomic nuclei. 

The details of an atomic system may be quite complicated. The exact
density distributions of the electrons are subject to both quantum
mechanical interference effects and complicated screening effects
which result from the electron-electron interaction. However, the
basic mechanical properties of size and compressibility may be 
explained without examining these details, just as it is done in
macroscopic systems. For this purpose, we need to define a kind of
mean field potential for the Coulomb interactions, including electron-
electron repulsion.

Effectively, there is one positive charge in the nucleus for each electron.
Therefore, we will estimate the Coulomb energy of a given electron density 
by assuming that the charge of each electron is evenly distributed in a 
sphere with volume $\rho^{-1}$ around a single positive charge.
The Coulomb energy per electron is then given by

\begin{equation}
E_{Coulomb}=-\frac{3 e^2}{8\pi\epsilon_0}(\frac{4\pi}{3}\rho)^{1/3}.  
\end{equation}

Note that by ignoring the electrons which are much closer to 
the nucleus than most others and therefore see the "naked" charge
of the nucleus we have essentially chosen to describe
only the valence electrons. In fact, the "nucleus" is effectively
an ion with a closed shell. The properties of size and elasticity 
are not affected by this ion, however, and we can therefore ignore
this hidden part of the electronic system in this context.
The choice of only a single positive charge for each electron 
should be best for alkali metals, so quantitative results should 
be tested by comparison with these elements.  

The complete thermodynamic relation for the energy  $E_{el}$ per
electron as a function of the volume $V_{el} = 
\rho^{-1}$ per electron is

\begin{equation}
E_{el}(V_{el})\approx \frac{3\hbar^2}{10m} (\frac{3\pi^2}{V_{el}})^{2/3}
                    -\frac{3e^2}{8\pi\epsilon_0} (\frac{4\pi}{3 V_{el}})^{1/3}. 
\end{equation}

The volume occupied by each electron in equilibrium can be determined 
by finding
the minimum of the energy $E_{el}$. This minimum is found by setting
the derivative $dE/dV$ to 0.
Note, that the volume derivative of the energy is a pressure. Effectively, 
we determine the point of equilibrium between the Pauli pressure and the
Coulomb pressure. In the atom, the electrons are compressed by the 
Coulomb attraction until the repulsion caused by the Pauli principle 
compensates this force. 

The length scale at which this happens is, of course, the atomic length 
scale of the Bohr radius $a_0$.

\begin{equation}
a_0=\frac{4\pi\epsilon_0\hbar^2}{me^2}= 0.529\times 10^{-10}m
\end{equation}

The volume per electron calculated from the relation given above 
is $V_{el}= 2 (3\pi^3 a_0/5)^3 \approx 4\pi a_0^3$, roughly equal to the 
volume of a sphere
with radius $1.5 a_0$. However, the approximation of the Coulomb pressure
used is quite rough, so that only the order of magnitude is meaningful.

To get an idea of the pressures involved, we can now calculate the 
Pauli pressure and the Coulomb pressure separately to find that for
$V_{el}\approx 4\pi a_0^3$, this preassure is around $10^12 Pa$ or 
10 million times atmospheric pressure. 

To get an idea of how "solid" the atom actually is, we can
estimate the compressibility $K$ of the atom, defined as

\begin{equation}
K=(V\frac{d^2 E}{dV^2})^{-1}.
\end{equation}

Not surprisingly, the result is
roughly equal to the inverse Pauli pressure. For $V_{el} \approx 4\pi a_0^3$,
we find $K \approx 5 \times 10^{-12} Pa^{-1}$.
This is exactly the order of magnitude found in the compressibilities 
of the alkali metals. For metals in general, typical values range from 
caesium at
$7 \times 10^{-12} Pa^{-1}$ to copper at 
$0.0745 \times 10^{-12} Pa^{-1}$.
Note that the compressibility of metals does not depend on the properties
of the metallic bond: we are not considering the effects which make atoms
stick together yet. To check, whether atoms which form only weak bonds show 
the same mechanical properties as the atoms of metals, a comparison with the 
noble gases is useful.
The solids formed by condensing the noble gases at very low temperatures
have compressibilities from $9\times 10^{-12} Pa^{-1}$ for neon to
$2.8 \times 10^{-12} Pa^{-1}$ for xenon. This is indeed very close to the
alkali metals. Another type of crystal in which the hard sphere model of
atoms seems to be appropriate is the ionic crystal. Although the charge
distribution in those crystals is very inhomogeneous, their compressibility
is only one order of magnitude less than that of the alkali metals and the
noble gases, with typical values close to the compressibility of
$0.42 \times 10^{-12} Pa^{-1}$ for sodium chloride. All values are taken
from the corresponding chapters of \cite{2}. 

The result that the compressibility of single atoms is close to
the compressibility of solids is well in tune with the quite
intuitive picture of solids consisting of densly packed atoms.
As a result of the immense magnitude of the Pauli pressure and the 
Coulomb pressure, the hard sphere model is quite an adequate picture for the
atom despite the misleading notion of emptiness associated with the assumed 
"sizes" of electrons and nuclei. 
We can therefore conclude that the source of the low compressibility 
of solids is indeed the Pauli pressure and that the atomic properties
defined by the equilibrium between Coulomb attraction and Pauli pressure
are not changed very much when chemical bonds are formed. 

However, the Pauli pressure and the Coulomb pressure may be compensated
atom by atom, and nothing in this picture makes the seperate hard sphere 
atoms stick together. This is approximately true for the noble gases,
but certainly not for most atoms. The periodic table of the elements
shows us that the reacivity of the melements depends on the exact number
of electrons.  
Therefore, we need to take a closer look at the particle properties 
of the electronic system in order to understand the forces that hold 
atoms together.

\section{Unpaired electrons}
Since electrons are particles with spin 1/2 and can therefore occupy two spin
states for every real space state, the Pauli pressure seems to act
seperately on two types of electrons. If the electron gas were continuous
and did not consist of quantized particles, the spin up density would thus
always equal the spin down density, and all matter would be paramagnetic.
Indeed, this type of paramagnetism exists in metals and is consequently
known as Pauli paramagnetism \cite{2}. 

However, in atoms there are a discrete number of electrons in a discrete 
number of states. Therefore, it may happen that n degenerate states are 
filled with less than
2n electrons. In this case, the Pauli preassure is insensitive to the
distribution of the electrons. Indeed, if all unpaired electrons align their
spins, the Pauli principle keeps them spatially separate, reducing the
electron-electron Coulomb repulsion, while electrons with opposite spins
would overlap more strongly, even if in separate real space states. This
energy is the spin exchange energy calculated in Hartree-Fock theory. If we
have a high degeneracy of states partially filled, we therefore have a ground
state with all unpaired spins parallel. This is the reason for both Hund`s
first rule and for ferromagnetism.

Energetically, the unpaired electrons are always somewhat unstable, as are 
all systems with a degenerate ground state, since any small perturbation will
remove the degeneracy and break the symmetry of the system by forming new 
geometric patterns. In surface science, this tendency for forming bonds is 
often referred to as Jahn-Teller instabitity \cite{4}. Although the change in
geometry is not as impressive in the liquid or gas phase, the reactivity of 
atoms with partially filled shells is a consequence of an instability of the 
same type.

The effect which causes the perturbation when another
similar atom approaches is not, however, the electric field, although
van-der-Waals forces will be present. Instead, homopolar bond formation 
is largely the effect of kinetic energy, as the nearly free electron bands 
in solid state physics, which are found even in non-metallic systems, 
suggest. The very small effects of charge density redistribution 
in a molecule are quite insufficient to explain the energy of chemical
bonds, especially since the energetically favorable positions are always
the positions close to the atomic nucleus and not the position between the
nuclei. In textbooks, this is often explained in a misleading way, 
because it seems so much harder to visualize the effects of kinetic
energy in a picture. If the properties of atoms that the kinetic energy 
and the potential energy are related by the equilibrium of Pauli pressure 
and Coulomb pressure are applied to the equations, it is even possible
to express the kinetic energy contribution to bonding in terms of the
potential energy of the atomic electrons. To avoid this type of confusion,
We will therefore neglect van-der-Waals forces and polar contribution 
altogether and determine the effect of 
kinetic energy only. 

This can also be justified by examining the Hamiltonian of the two atom
system:

\begin{equation}
\hat{H}=\hat{T}+\hat{V}_1+\hat{V}_2+\hat{V}_{el-el}
\end{equation}

$\hat{T}$ is the kinetic energy of all the electrons, $\hat{V}_{1/2}$ is the
potential of the respective atomic nucleus, and $\hat{V}_{el-el}$ is the
electron-electron interaction. Considering that due to a negligible overlap
of the electron distribution the Coulomb energies, including
electron-electron interactions, act almost exclusively on wavefunctions
localized at the respective atoms, this Hamiltonian may be seperated into
atomic Hamiltonians:
\begin{equation}
\hat{H}_1=\hat{T}+\hat{V}_1+\hat{V}_{el-el,1}
\end{equation}
\begin{equation}
\hat{H}_2=\hat{T}+\hat{V}_2+\hat{V}_{el-el,2}.
\end{equation}
The eigenstates of $\hat{H}_1$ and $\hat{H}_1$ are the electronic states of
the individual atoms. However, only the potential energy terms are localized 
at the site of the atom. The kinetic energy term $\hat{T}$ is the sum of all
electronic kinetic energies regardless of the atom at which the electron is
located. The total Hamiltonian is therefore given by the sum of the atomic
Hamiltonians minus once the kinetic energy term.
\begin{equation}
\label{eq:decomp}
\hat{H}=\hat{H}_1+\hat{H}_2-\hat{T}
\end{equation}
For electronic states located entirely at atom 1, the contributions of 
$\hat{H}_2-\hat{T}$ are negligibly small and so are contributions of 
$\hat{H}_1-\hat{T}$ for states located entirely at atom 2. For delocalized
electronic states however,
equation (\ref{eq:decomp}) is a useful expression
for the evaluation of energy contributions due to the delocalization
as we shall see below.

Of course, we have thus neglected the electron-electron interactions between
electrons located at different atoms, as well as interactions between an
atomic nucleus and electrons located at the other nucleus. For well 
separated atoms,
the net effect of these interactions are the van-der-Waals forces and ionic
interactions. We have neglected these effects, because van-der-Waals forces
are known to be much weaker than the forces involved in chemical bonding
and ionic effects are expected to be weak in the case of symmetric homopolar
bonds. To treat the case of polar bonds, the ionic interaction term must be 
considered in more detail.

Many particle effects within the atoms are still included in the Hamiltonian.
However, they do not contribute to the bond energy. 
The kinetic energy term responsible for the bond formation is 
generated because the kinetic energy is included
in both atomic Hamiltonians, and must therefore be subtracted once from the 
total. This indicates the fact, that the kinetic energy is not localized at 
one of the atoms, but causes a delocalization connecting the two atoms.

The kinetic energy is a single particle property. Therefore, we can 
continue by calculating the energy contributions for each electron 
separately, adding the energy terms of all electrons involved in
the bond formation to obtain the total energy. The single particle
wavefunctions may be obtained by Hartree-Fock theory, together with the
effective single particle Hamiltonian for each atom.

The matrix element connecting two electronic orbitals at
different atoms, $|1>$ and $|2>$, is given by \\[0.5cm]
\begin{math}
<1|\hat{H}^{eff}|2>=
\end{math}
\begin{equation}
<1|\hat{H}^{eff}_1|2>+<1|\hat{H}^{eff}_2|2>-<1|\hat{T}|2>,
\end{equation}
where $\hat{H}^{eff}_i$ is the effective single electron Hamiltonian
of atom $i$.
Since $|1>$ and $|2>$ must be eigenstates of the respective atomic 
Hamiltonians the
first two contributions are just the eigenenergies multiplied with the 
direct overlap $<1|2>$. The latter is a correction for the fact that, 
since $|1>$ and $|2>$ are
not orthogonal, a part of the matrix element represents the expectation
value of the Energy of a state localized at one of the two atoms. If the
non-orthogonality is removed, this contribution goes to 0, and the remaining
matrix element is purely kinetic:
\begin{equation}
<1|\hat{H}^{eff}|2>=-<1|\hat{T}|2>
\end{equation}
Note, that the sign already suggests a lower energy for the symmetric state!

It is now possible to develope a complete theory of bonding by defining the
atomic wavefunctions and their energies and calculating the tight binding
matrix elements using only the operator of the kinetic energy.

\section{Bonding and anti-bonding states}
If we consider only one electronic orbital per atom,
the bonding and anti-bonding states of a homopolar bond
must be the sum and the difference of the atomic wavefunctions. 
The energies of the bonding and anti-bonding states can then be 
determined by calculating the expectation values of the kinetic 
energy for these two linear combinations. Note, that the sum and the 
difference of two normalized wavefunctions are automatically orthogonal to 
each other. If the atomic wavefunctions are given by $\psi_{1/2}({\bf r})$,

\begin{mathletters}
\begin{equation}
\psi_\pm({\bf d})=\frac{1}{\sqrt{N_\pm}}
(\psi_1({\bf r}+{\bf d}/2)\pm\psi_2({\bf r}-{\bf d}/2))
\end{equation}
\begin{math}
N_\pm=2\pm\int(\psi_1({\bf r}+{\bf d}/2)\psi^*_2({\bf r}-{\bf d}/2)
\end{math}
\begin{equation}
              +\psi_1^*({\bf r}+{\bf d}/2)\psi_2({\bf r}-{\bf d}/2))d^3{\bf r}
\end{equation}
\end{mathletters}

\[
<E_{kin}>_{\pm}({\bf d})=\frac{1}{N_\pm}(<E_{kin}>_{1}+<E_{kin}>_{2}
\]\[
\mp\frac{\hbar^2}{2m}
\int (grad(\psi_1({\bf r}+{\bf d}/2)) grad(\psi^*_2({\bf r}-{\bf d}/2))
\]
\begin{equation}
+grad(\psi_1^*({\bf r}+{\bf d}/2)) grad(\psi_2({\bf r}-{\bf d}/2)))d^3{\bf r}
\end{equation}

If each atom contributes one unpaired electron, these two electrons
can redistribute into the two bonding and two anti-bonding levels as
the atoms approach each other. If they pair up in the bonding
level, a pair bond is formed.

Although these real space integrals already describe the effect of
wavefunction overlap on the kinetic energy completely, it is quite revealing
to take a look at the same equations in k space by Fourier transforming the
whole integral, which represents a convolution of two gradients. In k space,
this changes into the integral of the product of the wavefunctions multiplied
by $\bf{k}^2$ and a cosine which represents the effect of the real space
separation:
\[
<E_{kin}>_{\pm}({\bf d})=
\]
\[
\frac{1}{N_\pm}(<E_{kin}>_{1}+<E_{kin}>_{2}
\]
\[
\mp\frac{\hbar^2}{2m}
\int({\bf k}^2(\psi_1({\bf k})\psi^*_2({\bf k})e^{-i{\bf kd}}
\]
\begin{equation}
              +\psi_1^*({\bf k})\psi_2({bf k})e^{+i{bf kd}}))d^3{\bf k}
\end{equation}

If $\psi_1=\psi_2$, the kinetic energy contributions of the overlapping
wavefunctions can be written in an even more compact form:
\[
<E_{kin}>_{\pm}({\bf d})=
\]
\[
\frac{1}{N_\pm}\frac{\hbar^2}{2m}
\int(2{\bf k}^2\psi({\bf k})\psi^*({\bf k})(1\pm cos({\bf kd})))d^3{\bf k}
\]
\begin{equation}
=\frac{\hbar^2}{2m}
\frac{\int({\bf k}^2\psi({\bf k})\psi^*({\bf k})(1\pm cos({\bf kd})))d^3{\bf
k}}{\int(\psi({\bf k})\psi^*({\bf k})(1\pm cos({\bf kd})))d^3{\bf k}}
\end{equation}

In this case, the kinetic energy distribution is therefore modified by a
factor of $1\pm cos({\bf kd})$ at each point in k space.
If the width of the impulse distribution of the wavefunctions, that is, the
width in k space of $\psi^*\psi$, is roughly equal to $\pi/d$, then
the contributions with the highest $k^2$ values at the edge of the
distribution are suppressed by a multiplication with values close to 0 for the bonding state. The
kinetic energy of the bond is therefore at a minimum. If the k space width of
the atomic wavefunctions is much larger than $\pi/d$, the rapid oscillations
of the cosine make all overlap contributions cancel (as they should, since
the atoms are far apart in real space now). If the k space width is much
smaller
than $\pi/d$, all parts of the k-space distribution contribute equally and
the bonding state is again equivalent to the atomic wavefunction\cite{5}.

The k space distribution of the atomic wavefunction therefore defines a bond
length of roughly $\pi$ divided by its width in k-space, at which the average
wavelength of the electron is in resonance with the bond length. Since the
k-space width is connected with the real space width by the uncertainty
relation, this bond length is roughly equal to the sum of the atomic radii,
so the covalent bond is actually strongest when the atoms just about touch
each other.

To understand the significance of this result, it should be remembered, that
the potential energy effects only enter the picture indirectly, by forming
the atomic wavefunctions. The formation of the chemical bond can then be
explained entirely by the change of kinetic energy when unpaired electrons
begin to tunnel resonantly between the atoms. While the total electron
density in real space changes only little, the kinetic energy distribution
looses a major part of its high energy contributions.

\section{A quantitative example: Gaussian wavefunctions}
As an analytically solvable example, Gaussian wavefunctions offer a good
insight into the formation of bonds by kinetic energy resonance. The most
simple bond is a symmetric combination of s-type states:

\begin{mathletters}
\begin{equation}
\psi({\bf r})=\frac{1}{(2\pi\sigma^2)^{3/4}}e^{-{\bf r}^2/4\sigma^2}
\end{equation}

Since the Fouriertransform of the Gaussian is again a Gaussian, it is easily
possible to determine all features of the resulting bond between two unpaired
electrons in such atomic states.

\begin{equation}
\psi({\bf k})=(\frac{2\sigma^2}{\pi})^{3/4}e^{-\sigma^2{\bf k}^2}
\end{equation}
\begin{equation}
<E_{kin}>=\frac{3\hbar^2}{4m\sigma^2}
\end{equation}
\end{mathletters}

Note that the factor of three is a consequence of the three spatial
dimensions contributing to the energy. Since we can separate the dimensions
for Gaussian wavefunctions, we could as well examine only the one dimensional
problem. However, we shall include the constant energy contributions for the
sake of completeness.

\begin{mathletters}
\begin{equation}
\psi_+({\bf k})=\sqrt{\frac{2}{1+e^{-1/2(d/2\sigma)^2}}} \; cos({\bf dk}/2)
(\frac{2\sigma^2}{\pi})^{3/4}e^{-\sigma^2{\bf k}^2}
\end{equation}

\[
<E_{kin}>=\frac{\hbar^2}{4m\sigma^2}
\frac{3+(3-(d/2\sigma)^2)exp(-1/2(d/2\sigma)^2)}{1+exp(-1/2(d/2\sigma)^2)}
\]
\begin{equation}
=\frac{3\hbar^2}{4m\sigma^2}-\frac{\hbar^2}{4m\sigma^2}
\frac{(d/2\sigma)^2}{exp(+1/2(d/2\sigma)^2)+1}
\end{equation}
\end{mathletters}

\begin{mathletters}
\begin{equation}
\psi_-({\bf k})=i\sqrt{\frac{2}{1-e^{-1/2(d/2\sigma)^2}}} \; sin({\bf dk}/2)
(\frac{2\sigma^2}{\pi})^{3/4}e^{-\sigma^2{\bf k}^2}
\end{equation}
\[<E_{kin}>=\frac{\hbar^2}{4m\sigma^2}
\frac{3-(3-(d/2\sigma)^2)exp(-1/2(d/2\sigma)^2)}{1-exp(-1/2(d/2\sigma)^2)}
\]
\begin{equation}
=\frac{3\hbar^2}{4m\sigma^2}+\frac{\hbar^2}{4m\sigma^2}
\frac{(d/2\sigma)^2}{exp(+1/2(d/2\sigma)^2)-1}
\end{equation}
\end{mathletters}

Figure 1 shows the energy and the k-space distribution of the bonding state as a function of d and figure 2 shows the same for the anti-bonding state.\\
We can now determine the bond length, as well as the harmonic part of the
potential around it. This allows an estimate of typical molecular vibrations
and a comparison with experimental values. 
 
The minimum of the bonding potential is at $d=3.2 \sigma$. This corresponds very well with the estimate given in the previous section, that the bond length should be roughly equal to $\pi$ divided by the width of the distribution in k-space. The standard deviation of the k-space distribution is $1/2\sigma$, so $1/\sigma$ is a good measure of its width. \\
Around the minimum at $d=3.2 \sigma$, the bonding potential may be written as
\begin{equation}
<E_{kin}>=\frac{\hbar^2}{4m\sigma^2}
((3-0.557)+0.55(d/2\sigma-1.6)^2).
\end{equation}
The total energy is then given by Z times this potential, where Z is the
number of electrons in the bond. From this potential, the following relations
for bondlength $d_0$, bond energy $E_b$ and the bond elasticity k can be
obtained:

\begin{mathletters}
\begin{equation}
d_0=3.2\sigma
\end{equation}

\begin{equation}
E_b=51 (\frac{10^{-10}m}{\sigma})^2 Z \; [\frac{kJ}{mol}]
\end{equation}

\begin{equation}
k=4.2 (\frac{10^{-10}m}{\sigma})^4 Z  \; [\frac{N}{m}]
\end{equation}
\end{mathletters}

Table 1 lists a few examples of diatomic molecules with their bond length,
bonding energy and the elastic constant. Data for the elastic constant was 
taken from \cite{6} and all other data from \cite{1}. The table also shows 
the $\sigma$ values that would correspond to the respective bond properties 
if the wavefunctions were simple
Gaussians. Although there are strong deviations even between different
properties of the same molecule, the order of magnitude is reproduced
correctly despite the simplicity of the model.\\
To simulate a directional bond, we can calculate the bonding potential for 
two  wavefunctions of p-type symmetry:

\begin{mathletters}
\begin{equation}
\psi({\bf r})=\frac{x}{\sigma(2\pi\sigma^2)^{3/4}}e^{-{\bf r}^2/4\sigma^2}
\end{equation}
\begin{equation}
\psi({\bf k})=2\sigma k_x (\frac{2\sigma^2}{\pi})^{3/4}e^{-\sigma^2{\bf k}^2}
\end{equation}
\begin{equation}
<E_{kin}>=\frac{5\hbar^2}{4m\sigma^2}
\end{equation}
\end{mathletters}

Note, that the separation between the two atoms, {\bf d}, need not be along 
the x axis. It is possible to calculate the potential for the additive and 
the subtractive linear combinations as before. The potential for the bonding 
state in three dimensions is given by
\[
<E_{kin}>_-=\frac{5\hbar^2}{4m\sigma^2}+
\]
\[
\frac{\hbar^2}{4m\sigma^2}
(\frac{(d_x/2\sigma)^2(({\bf d}/2\sigma)^2-3}{1-({\bf d}/2\sigma)^2
-exp(+1/2({\bf d}/2\sigma)^2)}
\]
\begin{equation}
+\frac{(d_y/2\sigma)^2+(d_z/2\sigma)^2}
{exp(+1/2({\bf d}/2\sigma)^2)-1})
\end{equation}

This equation describes a potential with two minima along the x axis, at 
$d_x=\pm 5.04\sigma$, as is shown in figure 3. It represents the bonding 
potential of a $pp\sigma$ bond, since there is rotational symmetry around 
the x-axis. In the vicinity of these minima, the potential may be 
approximated by a harmonic potential of

\[
<E_{kin}>=\frac{\hbar^2}{4m\sigma^2}
\]
\begin{equation}
(5-0.73+1.08(\frac{d_x}{2\sigma}-2.52)^2+0.42((\frac{d_y}{2\sigma})^2
+(\frac{d_z}{2\sigma})^2)).
\end{equation}

It is now possible to compare the longitudinal and the transversal
elasticity of the bond. The ratio of the corresponding coefficients is 
$1.08/0.42$. The bond is therefore softer against shear forces than against 
compression by a factor of about 2.5. This example shows how the geometrical 
structure of molecules and their vibrations can be described using only the 
kinetic energy and the symmetry of atomic wavefunctions. Of course, a more 
comprehensive description of molecules or crystals would require the full 
calculation of matrix elements between all the atomic wavefunctions involved,
using more realistic wavefunctions than the Gaussians presented here.

\section{The ionic contribution: extending the model to polar bonds}
The kinetic energy only dominates the homopolar bond. In the case of 
heteropolar bonds, there will be an energy difference between the electronic 
states at atom A and at atom B, which pulls the electron towards the more 
electronegative one. In a very simpleminded tight binding approach, this 
effect may be included as a site dependend energy. With $E_A-E_B=2D$, the 
two by two matrix of the polar bond is then given by
\begin{equation}
\left(
\begin{array}{cc}
T_+ & D \\
D & T_- 
\end{array}
\right)
\end{equation}
$T_\pm$ is the kinetic energy of the bonding and anti-bonding states. For 
large separations d, this is an unrealistic approach, however, since it 
neglects the energy needed to ionize the atoms and therefore predicts 
complete ionization as d approaches infinity, not including the Coulomb 
attraction.\\
The other extreme is to assume fully polarized atoms. In this case, a term 
representing the Pauli repulsion is necessary to keep the ions apart. This 
may be done by adding the bonding and anti-bonding energies for the filled 
outer shells of the ions.\\To combine covalent and ionic effects into a 
model of the polar bond, it is necessary to consider a total of three 
possible electronic configurations:  state 1, with both electrons at atom 
A, state 2 with one electron at each atom, and state 3 with both electrons 
at atom B. Note that this corresponds to a simple extension  of 
Heitler-London theory \cite{6}. The Hamiltonian matrix is 
\begin{equation}
\left(
\begin{array}{ccc}
T_++T_-+V_A & \sqrt{1/2}(T_+-T_-) & 0 \\
\sqrt{1/2}(T_+-T_-) & T_++T_- & \sqrt{1/2}(T_+-T_-) \\
0 & \sqrt{1/2}(T_+-T_-) & T_++T_-+V_B
\end{array}
\right)
\end{equation}
For $V_A=+D$ and $V_B=-D$, this is equivalent to the two state model above. 
However, by including the ionization energy in the $V_{A/B}$, we can now 
correct the result for large d. For very large d, $V_{A/B}$ is equal to the 
total ionization energies $V_{A/B}(\infty)$ required to remove an electron 
from one atom and add it to the other. V then follows the Coulomb law until 
the atoms come quite close, when the atomic wavefunctions start to penetrate 
each other. At very small distances, the difference in energy is given by 
the energy level difference of the atomic wavefunctions, $D=1/2(E_A-E_B)$. 
The transition from the Coulomb regime to the energy level difference regime 
may be extrapolated using a function of the following form:
n
\begin{mathletters}
\begin{equation}
V_{A/B}(d)= V_{A/B}(\infty)+\frac{\pm D - V_{A/B}(\infty)}
{\sqrt{1+(d/d_{A/B})^2}}
\end{equation}
\begin{equation}
d_{A/B}=\frac{e^2}{4\pi\epsilon_0 (\pm D - V_{A/B}(\infty)}
\end{equation}
\end{mathletters}

The length $d_{A/B}$ defines the length scale at which the transition 
between covalent and polar bonding occurs. Since this length is largely 
determined by the atomic size, it is typically close to the bond length. 
Still, the effect of $V_{A/B}(\infty)$ can often be ignored, since the 
effects of the energy level difference and the covalent bond combined 
tend to be stronger at atomic separations equal to the bond length. 

\section{Conclusions}
While many textbook explanations of chemical bonds try to visualize the bond 
only in its spacial distribution, misleadingly suggesting that the source of 
the bond energy could be the slight increase in electron density between the 
atoms, the approach presented here clearly identifies the kinetic energy 
term as the dominant contribution to homopolar bond formation. In the spirit 
of the tight binding approximation, one can consider the atomic confinement 
of the electrons as much stronger than the interatomic interactions, and 
thereby arrive at a bonding potential by simply calculating the expectation 
values of the kinetic energy for electron pairs evenly distributed between 
the atoms. This type of bonding potential, which uses only the unchanged 
atomic wavefunction and the fundamental kinetic energy term of free 
electrons, not only gives the right order of magnitude for the bond energy,
but also reproduces the spatial potential of the bond, with realistic results for bond length and bond elasticity. 
 
Although a quantitatively accurate calculation may only be achieved by 
including a more detailed description of polar effects than discussed here, 
it should be pointed out once more, that purely kinetic matrix elements do 
give highly accurate results in semiconductors and other simple crystals. 
Furthermore, the free electron behaviour of electrons in metals and the 
relation between covalent and metallic bonds may be understood better in the 
light of these considerations. In fact, the major difference between the two 
type of bonds is not the mechanism of bonding itself, but rather concerns 
the fact  that metallic bonds are not directed as are bonds involving p type 
wavefunctions. Instead they couple equally well to all neighbours, resulting 
in the strong delocalization of electrons which makes metals conductors and 
in their relatively high plasticity compared with the brittle covalent bonds 
of e.g. semiconductors. 

In this manner, it is possible to find numerous connections between the very 
fundamental laws of quantum physics and the chemical and physical properties 
of the world surrounding us. Indeed, as technology advances it is  
important to remember that science not only tells us how to do things, but 
also why things are as they are, although this often teaches us more 
about limitations than about possibilities. Even the best artificial 
materials will not be orders of magnitudes removed from the typical 
properties dictated by constants such as $\hbar$ and $e$, just as no signal 
will ever travel faster than the speed of light. However, knowledge of our 
limitations may often proof more important than know-how, and in this sense, 
simplified models as the one presented here can proof to be extremely useful.

\section*{Acknowledgements}
I would like to thank Dr. O. Hess for encouraging me to write this article
 and Ulf Willhelm and Stefan Rodewald for taking an active interest in it.

\newpage
\onecolumn
\underline{Table 1:}
\begin{center}
\begin{tabular}{|c|c|cc|cc|cc|}
& & bond & & bond & & bond &\\
Molecule & Z & length & $\sigma$  
& energy & $\sigma$ & elasticity & 
$\sigma$\\
& & in $10^{-10}m$ &in $10^{-10}m$ & in kJ/mol &in $10^{-10}m$ & in N/m &
in $10^{-10}m$ \\ \hline
$H_2 $ & 2 & 0.74 & 0.23 & 432 & 0.49 & 520 & 0.36 \\
$Li_2$ & 2 & 2.67 & 0.83 & 110 & 0.96 & 130 & 0.50 \\
$Na_2$ & 2 & 3.08 & 0.96 & 72  & 1.19 & 170 & 0.47 \\
$N_2 $ & 6 & 1.10 & 0.34 & 843 & 0.60 & 2260& 0.32 \\
$O_2 $ & 4 & 1.20 & 0.38 & 494 & 0.64 & 1140& 0.35 \\
$F_2 $ & 2 & 1.42 & 0.44 & 140 & 0.85 & 450 & 0.37 \\
$Cl_2$ & 2 & 2.00 & 0.63 & 240 & 0.65 & 320 & 0.40 \\
$Br_2$ & 2 & 2.28 & 0.71 & 190 & 0.73 & 240 & 0.43 \\
$CO  $ & 4 & 1.13 & 0.35 &1071 & 0.44 & 187 & 0.55 \\
$NO  $ & 4 & 1.15 & 0.36 & 678 & 0.55 & 155 & 0.57 \\
$HCl $ & 2 & 1.27 & 0.40 & 428 & 0.49 & 48  & 0.65 
\end{tabular}  
\\[1cm]
\end{center}
Table 1 shows the experimentally determined properties of some biatomic 
molecules, together with the value of the parameter $\sigma$ which would 
reproduce this property using Gaussian wavefunctions. Note that despite
the simplicity of the model, all values of $\sigma$ are of the same
order of magnitude.
 
\newpage
\twocolumn

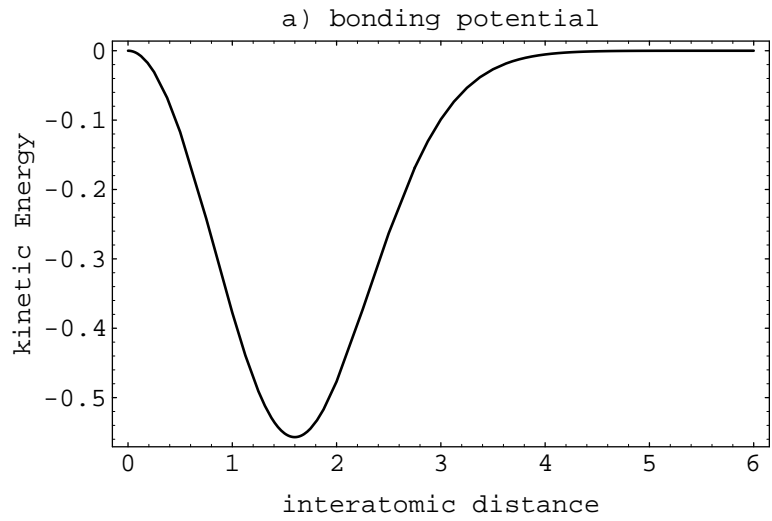
\underline{Figure 1:}

Figure 1a) shows the change in kinetic energy as a function of bond length 
for the bonding state.
The interatomic distance d is given in units of 2$\sigma$ and the energy 
difference from the unbound state is given in units of $\hbar/4m\sigma^2$. 

Figure 1b) shows the corresponding k-space distributions $|\psi(k)^2|$ as
a contour plot. the momentum $\hbar k$ is given in units of $\hbar/\sigma$. 

The minimum in bond energy occurs when the k-space distribution is most 
narrow, just before the side maxima appear. \\[0.5cm]
\underline{Figure 2:}

Figure 2a) shows the change in kinetic energy as a function of bond length
for the anti-bonding state.
all units are as in Figure 1.

For the anti-bonding state, the k-space distribution shows two peaks at small
distances d. It is therefore much wider than the bonding state distribution.
As d increases, it narrows, even though side maxima appear. \\[0.5cm]
\underline{Figure 3:}

Figure 3 shows the bonding potential of the $pp\sigma$ bond calculated 
for the p symmetry Gaussian wavefunctions in the xy plane. The unit of length
is $2\sigma$. 


\begin{thebibliography}{99}
\bibitem{1} {\it Chemical Principles}, R.E. Dickerson, H.B. Grey, M.Y. Darensbourgh and D.J. Darensbourgh, Benjamin Cummings Publishing 1984 
\bibitem{2} {\it Solid State Physics}, N.W. Ashcroft and N.D. Mermin, Saunders College Publishing 1976
\bibitem{3} Froyen and Harrison, PRB 20, 2420 (1979)
\bibitem{4} see for example {\it Physics of Surfaces}, A. Zangwill, Cambridge University Press, page 97, original publication: H.A. Jahn and E. Teller, Proc. Roy. Soc. A161, 220 (1937)
\bibitem{5} Note, that the anti-bonding state for small d is roughly k times
the atomic wavefunction if both atoms are similar. This wavefunction is
therefore similar to an excited atomic wavefunction, with the kinetic energy
increased accordingly.
\bibitem{6} {\it Molek\"ulphysik und Quantenchemie}, H. Haken and H.C. Wolf, Springer 1992
\end{thebibliography}
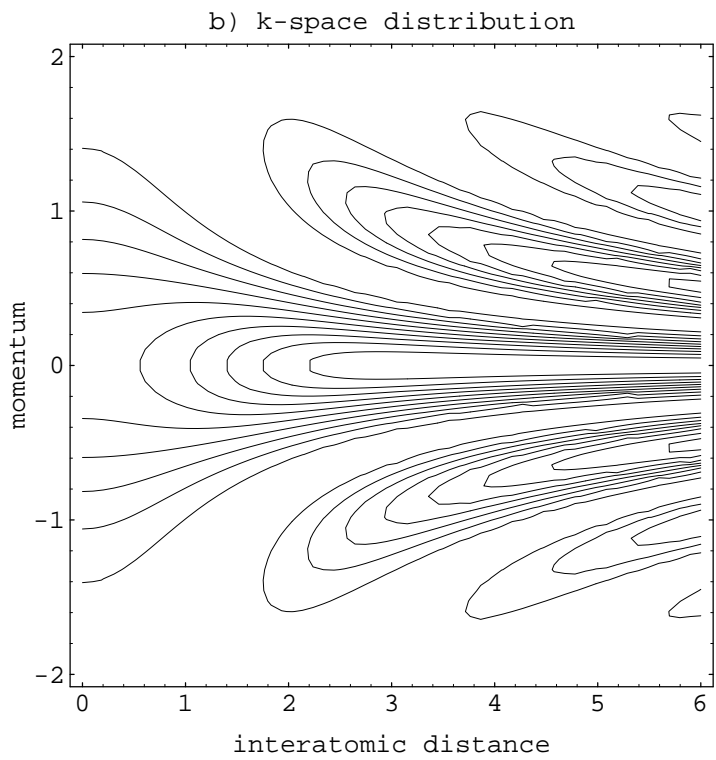
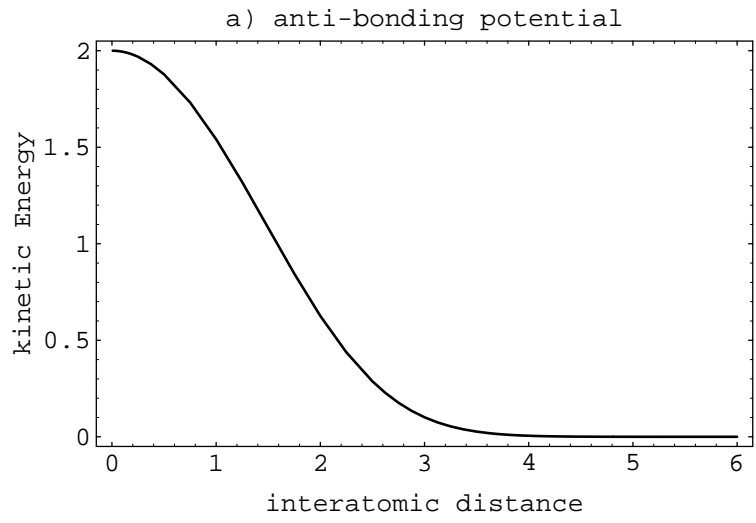
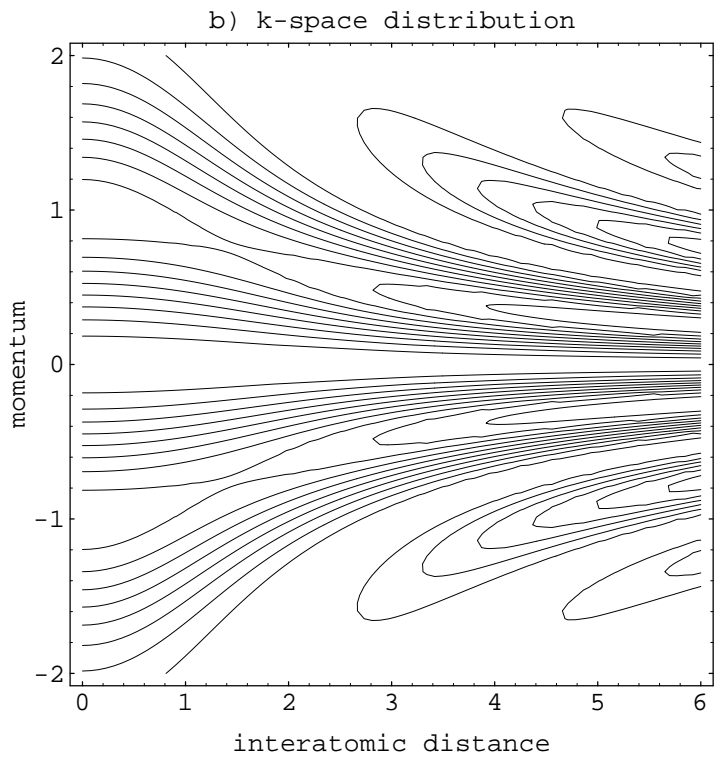
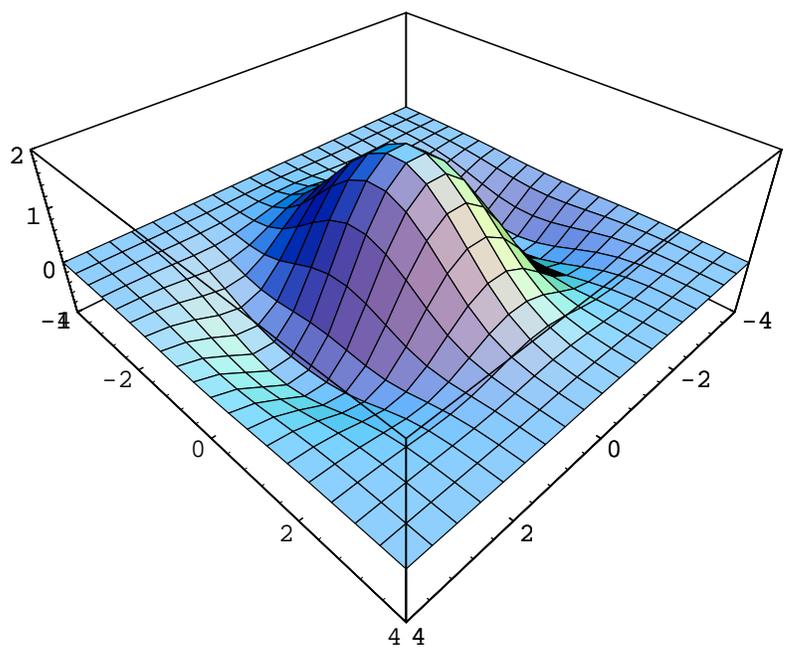
\end{document}